\documentstyle[epsf,psfig]{mn}
\begin{document}
\title[Formation times and environment]
      {On the environmental dependence of halo formation}
\author[R. K. Sheth \& G. Tormen]{Ravi K. Sheth$^1$ \& Giuseppe Tormen$^2$\\
$^1$ Department of Physics \& Astronomy, University of Pittsburgh, 
     3941 O'Hara Street, PA 15260, USA\\
$^2$ Dipartimento di Astronomia, Vicolo dell'Osservatorio 2, 
     35122 Padova, Italy \\
\smallskip
Email: rks12@pitt.edu, tormen@pd.astro.it
}
\date{Submitted to MNRAS 12 November 2003}

\maketitle

\begin{abstract}
A generic prediction of hierarchical gravitational clustering models 
is that the distribution of halo formation times should depend relatively 
strongly on halo mass, massive haloes forming more recently, and 
depend only weakly, if at all, on the large scale environment of the 
haloes.  We present a novel test of this assumption which uses the 
statistics of weighted or `marked' correlations, which prove to be 
particularly well-suited to detecting and quantifying weak 
correlations with environment.  We find that close pairs of haloes 
form at slightly higher redshifts than do more widely separated halo 
pairs, suggesting that haloes in dense regions form at slightly 
earlier times than do haloes of the same mass in less dense regions.  
The environmental trends we find are useful for models which relate 
the properties of galaxies to the formation histories of the haloes 
which surround them.  
\end{abstract}


\section{Introduction}
The excursion set model of hierarchical clustering (Epstein 1983; 
Bond et al. 1991) has been remarkably successful.  It provides 
useful analytic approximations for the abundance of haloes of mass 
$m$ at time $t$ (Bond et al. 1991; Sheth, Mo \& Tormen 2001), 
for the conditional mass function of $m$ haloes at $t$ which are 
later (at $T>t$) in more massive haloes $M>m$ (Bond et al. 1991; 
Lacey \& Cole 1993; Sheth \& Tormen 2002), for the abundance of 
haloes as a function of the larger scale environment 
(Mo \& White 1996; Lemson \& Kauffmann 1999; Sheth \& Tormen 2002)
for the distribution of halo formation times (Lacey \& Cole 1993) 
and masses (Nusser \& Sheth 1999; Sheth \& Tormen 2004).  Here, 
formation is typically defined as that time when the most massive 
progenitor contains at least half the final mass.  

In the simplest and most used approximation, this approach ignores 
most correlations between different spatial scales.  In this 
approximation, the approach predicts that there should be no correlation 
between halo formation and the large scale environment in which the 
halo sits (White 1996).  This is because, in the model, formation 
refers to a smaller mass than the final virial mass, and hence to a 
smaller spatial scale than that associated with the Lagrangian radius 
of an object, whereas the larger scale environment, by definition, 
refers to scales which are larger than that of the halo.  

Lemson \& Kauffmann (1999) presented evidence from measurements 
in numerical simulations of clustering that halo formation times 
were indeed independent of environment.  They interpreted this 
as evidence that the excursion set neglect of correlations was 
justified. 
(Lemson \& Kauffmann also presented evidence that a number of 
other physical properties of haloes were also independent of 
environment, and this evidence has been used to justify an 
assumption which enormously simplifies semi-analytic models of 
galaxy formation: that the properties of galaxies are determined 
by the haloes in which they form, and not by the surrounding 
larger-scale environment.)  
Their conclusion is somewhat surprising for the following reason.  
It is quite well established that the ratio of massive to low mass 
haloes is larger in dense regions, and that the excursion set model 
is able to quantify this dependence quite well (see the references 
given earlier).  
It is also well established that, on average, low mass haloes form 
at higher redshifts (see references given earlier).  
Together, these suggest that if one averages over the entire range 
of halo masses in any given region, then the mean formation time in 
dense regions should be shifted to lower redshifts, simply because 
these regions contain more massive haloes which, on average, form 
later.  In Figure~4 of their paper, Lemson \& Kauffmann averaged 
over the entire range of halo masses accessible to them in their 
simulations, and found no dependence of formation time on 
environement; at face value, this is {\it in}consistent with the 
simplest excursion set prediction!  

The main goal of this paper is to repeat the test for environmental 
effects on halo formation.  Section~\ref{density} shows that a 
simple plot of formation time versus local density does not show 
strong trends, suggesting that the excursion set approximation is 
rather accurate.  
But then, Section~\ref{marks} presents evidence, from what we feel is 
a more senstive test, which indicates that low mass haloes in dense 
environments form slightly earlier than haloes of the same mass in less 
dense environments.  Thus, it may be that, when one averages over a 
range of halo masses, the shift to later formation times associated 
with the fact that the dense regions contain the most massive haloes 
is approximately compensated-for by a density-dependent shift to 
slightly earlier formation times, with haloes in denser regions having 
larger formation redshifts than their counterparts (of the same mass) 
in the field.  This second test uses a technique known as marked 
correlation functions:  our results indicate that marked correlation 
functions are a powerful means of detecting and quantifying 
environmental dependences.  Section~\ref{alt} illustrates 
that our results do not depend sensitively on what definition of halo 
formation we use.  A final section summarizes our findings, and argues 
that our results may have important implications for studies of halo 
structure.  It also presents evidence which suggests that the density 
profiles of close halo pairs are neither more nor less centrally 
concentrated than are the profiles of their counterparts in less 
dense regions.

\section{Formation times and environment}\label{density}

\begin{figure}
\centering
\epsfxsize=\hsize\epsffile{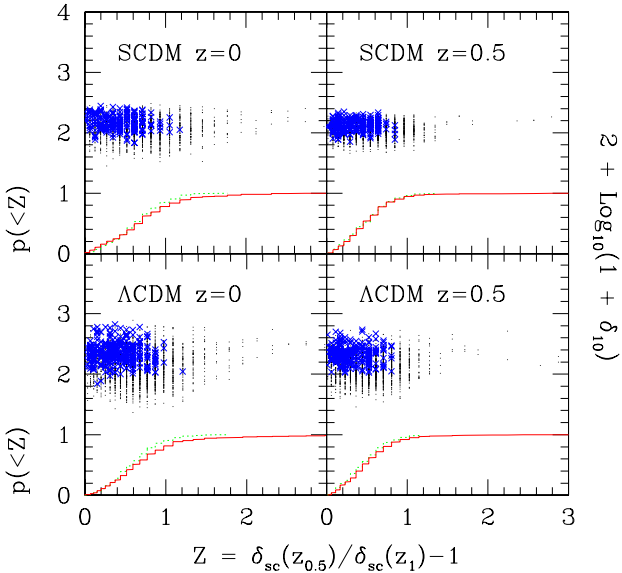}
\caption{Joint distribution of halo formation times and 
environments.  Crosses and dots represent massive and low mass 
haloes respectively.  Histograms show the cumulative distribution 
of formation redshifts, averaged over the entire halo population, 
for two bins in density (dotted/solid curves show results for 
low/high densities).  Although the symbols appear to show that 
low mass haloes in dense regions tend to have formation times which 
extend to higher redshifts, when averaged over the entire population 
of high- and low-mass haloes, there are no significant trends with 
environment. }
\label{zden}
\end{figure}

\begin{figure}
\centering
\epsfxsize=\hsize\epsffile{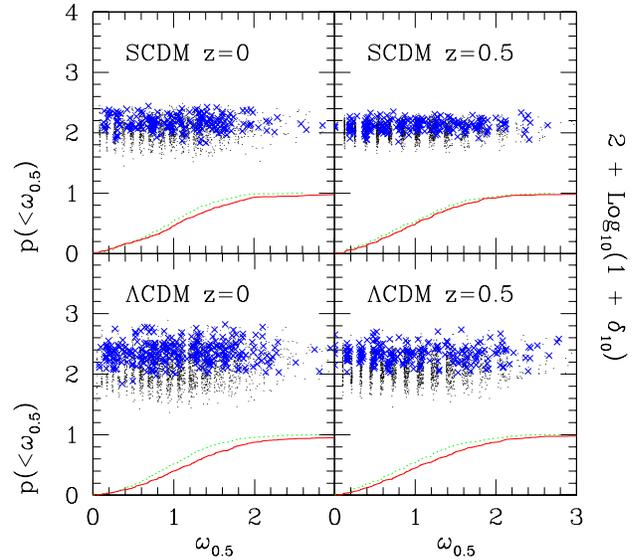}
\caption{Same as previous figure, but now with scaled formation 
redshifts.  Note that, in these scaled units, all haloes span the 
same range along the x-axis, whatever their mass.  
Histograms show the cumulative distribution of $\omega_{0.5}$, 
averaged over the entire halo population, for different bins in 
density (dotted/solid line is for lower/higher density environments).  
Comparison of the dotted and solid lines indicates that haloes in 
dense regions tend to have formation times which extend to higher 
redshifts compared to their counterparts in less dense regions.  }
\label{wden}
\end{figure}

We have identified the formation times of all haloes which contain 
more than two hundred particles in the GIF simulations of 
Kauffmann et al. (1999).  These simulations are a subset of those 
made available to the public by the Virgo consortium (Frenk et al. 2000).  
Particle positions and velocities from the simulations were output 
at a range of redshifts, approximately evenly spaced in logarithmic 
expansion factor: $\Delta\ln(1+z) \approx 0.0596$.  For each output 
time, we identified haloes using the spherical overdensity method 
(e.g. Lacey \& Cole 1994; Tormen, Moscardini \& Yoshida 2003)
which contained at least twenty particles.  The required overdensity 
is a cosmology dependent factor times the background density, as 
specified by the spherical collapse model.  For the SCDM model, this 
factor is 178, and it is independent of redshift; for the $\Lambda$CDM 
model, it is 323 at $z=0$, and is smaller at higher redshifts 
(e.g. Peebles 1993).  At any given output time $z_1$, we selected the 
halos which were composed of more than two hundred particles, and 
studied the formation times and masses at formation of these haloes as 
follows.  (For reference, an $M_*$ halo at $z=0$, 0.5 and 1.0 has 
1289, 170 and 31 particles in the SCDM run, and 807, 185 and 40 
particles in the $\Lambda$CDM run, so the high redshift runs mainly 
probe the formation times and masses of objects much larger than $M_*$.)  

Given a halo of mass $M_1$ (i.e., containing $N_1$ particles) at $z_1$, 
we go to the previous output time ($z_1$+d$z_2$, say), identify the 
object which contributes the most number of particles to $N_1$, and 
call it the most massive progenitor at $z_1+{\rm d}z_2$.  Suppose this 
most massive progenitor had $N_2$ particles.  We then go to the 
preceding output step ($z_1+ {\rm d}z_2+{\rm d}z_3$, say) and identify 
the most massive progenitor, $N_3$, of $N_2$.  We continue in this way 
until the number of particles in the most massive progenitor first 
falls below $N_1/2$.  If the mass just before formation is $N_n$, then 
the mass just after formation is $N_{n-1}$, and the redshift of formation 
is $z_1+\cdots+ {\rm d}z_{n-1}$.  We store these values for each halo 
$M_1$ at $z_1$.  

In what follows, we use $z_1$ 
to denote the redshift at which the parent object is identified as a 
virialized halo, and $z_{0.5}>z_1$ to denote the redshift at which 
it formed (recall formation is when the object is one-half its 
final mass).  
To estimate the local environment of a halo, we have measured the 
mass in spheres of radius $R=5$ and $10h^{-1}$~Mpc centred on the 
halo.  We then define the local overdensity on scale $R$ as 
 $\delta_{\rm R} = M_{\rm R}/(\bar\rho\,4\pi R^3/3) - 1$.  
We find qualitatively similar results for both $\delta_5$ and 
$\delta_{10}$ (the main difference being that 
$\delta_5\sim 10\delta_{10}$),  so in what follows, we only show 
results for $\delta_{10}$.  

Figure~\ref{zden} shows the joint distribution of halo formation 
redshifts and local densities $\delta_{\rm 10}$ (for clarity, we 
have shifted the estimate of the density upwards by 2 orders of 
magnitude, as indicated by the axis label on the right).  
The top panels show measurements for parent haloes identified in an 
SCDM simulation at $z=0$ (left) and $z=0.5$ (right), and the bottom 
panels show measurements in a $\Lambda$CDM simulation.  
In all panels, dots show results for low mass haloes (the subset 
with $0.25<M/M_*<0.5$ and $1<M/M_*<2$ for $z_1=0$ and 0.5, 
respectively), and crosses represent massive haloes 
($M/M_*>4$ and $M/M_*>16$ for  $z_1=0$ and 0.5).  

\begin{figure}
\centering
\epsfxsize=\hsize\epsffile{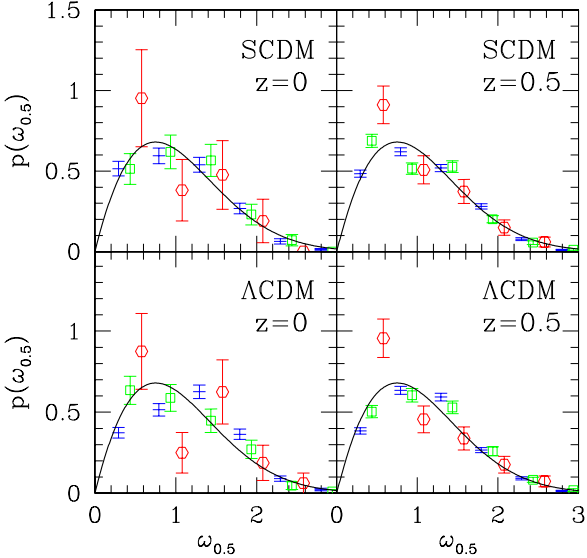}
\caption{Distribution of scaled formation times.  Panels on the left 
 and right show results for parent halos identified at $z=0$ and 
 $z=0.5$, and the different symbols in each panel show the 
 distribution of scaled formation times for halos with masses in 
 the range $1<M/M_*(z)\le 2$ (horizontal bars),
 $4<M/M_*(z)\le 8$ (squares),and $16<M/M_*(z)\le 32$ (hexagons).  
 Solid line shows the expression derived by Lacey \& Cole (1993).}
\label{pwlc}
\end{figure}

The figure shows clearly that the most massive haloes do not populate 
the least dense cells, and that they do not form at early times.  
The dotted and solid histograms in the bottom of each panel show the 
cumulative distribution of formation times, averaged over all halo 
masses (i.e., not just the ones shown by the dots and crosses), 
in low and high density regions.  
This was done by first sorting all haloes by their local density, 
and then choosing those whose local density was in the lowest ten 
percent to make the dotted curves, and haloes whose local density 
was in the highest ten percent to make the solid curves.  
The similarity of the cumulative distributions suggests that there 
is little evidence for dependence of formation time on local density.  
This is essentially what Lemson \& Kauffmann (1999) found, using a 
similar technique.  (In fact, Lemson \& Kauffmann defined the 
environment of a halo slightly differently---they use the density in 
a shell within 2$h^{-1}$ and 5$h^{-1}$Mpc of each halo, whereas we 
include the central region as well.  Since most halos are smaller 
than 2$h^{-1}$Mpc, our definition of local density places massive 
haloes in denser regions than does theirs.)  
Note, however, that as a result of the average over masses, this 
plot is difficult to interpret.  

A slightly more straightforward plot to interpret is shown in 
Figure~\ref{wden}, were we have scaled all the formation times to 
 $\omega_{0.5} = [\delta_{\rm sc}(z_{0.5})-\delta_{\rm sc}(z_1)]/
                         [\sigma^2(M/2)-\sigma^2(M)]^{1/2}$; 
in this variable, the dependence of formation time on mass has been 
removed (Lacey \& Cole 1993).  Comparison with the previous figure 
shows that, indeed, haloes of all masses now span the same range 
along the x-axis.  Figure~\ref{pwlc} shows this explicitly:  the 
different symbols show the scaled formation time distribution for 
different halo populations in the simulations.  The solid curves in 
each panel all show the same functional form, 
 $p(\omega) = 2\omega\,{\rm erfc}(\omega/\sqrt{2})$.  
Notice that this functional form provides a reasonably good 
description of the scaled halo formation times over a wide range of 
masses and times.  This rescaling allows us to look for a dependence 
on environment in the following sense: do haloes of a given mass in 
dense regions form at different times relative to their counterparts 
of the same mass in the field?  The slight separation between the 
cumulative histograms indicates that there is marginal evidence for 
higher formation redshifts in dense regions.  

Rather than quantifying this with a KS test, we have chosen to 
present the results of a different test which we believe is to 
be preferred, as it does not depend on the vagueries of what 
scale one chooses to define the local density (what is special 
about $10h^{-1}$Mpc? why make a spherical average?  what if one 
excises the region occupied by the halo itself when estimating 
the local density?), 
or on how the choice of density threshold affects the result 
(how do the cumulative histograms change if we simply split 
the sample into two equal-sized halves, rather than compare 
only the extreme ends of the density distribution?).  

\section{A marked correlation function analysis}\label{marks}
This section describes the results of a novel test of environmental 
effects on halo formation times:  an analysis of the `marked' 
correlations of the halo population, with formation time as the 
`mark'.  Let $m_i$ and $m_j$ denote the values of the marks 
associated with objects $i$ and $j$, and let $\bar m$ denote the 
mean value of the mark, when averaged over all objects.  
The marked correlation function we will use, $M(r)$, is defined as 
the sum over all pairs with separations $r_{ij}=r$, weighted by the 
product of the marks $m_im_j$, divided by the sum over the same pairs 
(i.e., those separated by $r$), but this time weighted by $\bar m^2$.  
In essence, $M(r)$ tests if pairs separated by $r$ tend to have 
larger or smaller values than the mean mark.  If we use the formation
redshift as the mark, then a plot of $M(r)$ versus $r$ shows if close 
pairs tend to have smaller ($M(r)<1$) or larger ($M(r)>1$) formation 
redshifts than average.  

\begin{figure}
\centering
\epsfxsize=\hsize\epsffile{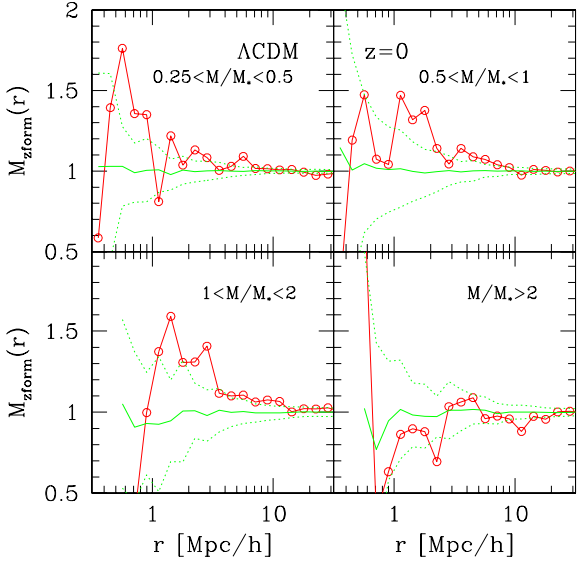}
\caption{Marked correlation function, with formation time as the 
 mark, for haloes identified at $z=0$ in the $\Lambda$CDM simulation. 
 Different panels show results for haloes of different masses.  
 Symbols show the measurement, and dotted line shows the typical 
 variation from the mean, shown as the heavy solid line, estimated 
 by computing $M(r)$ from a randomized set of marks, and repeating 
 one hundred times.  }
\label{xizf}
\end{figure}

The previous section showed that halo formation times depend on mass.  
Since the clustering of haloes depends on their mass, an analysis 
which averages over a large range in halo masses will be difficult 
to interpret.  Therefore, the different panels in Figure~\ref{xizf} 
show the marked correlation functions for a number of small bins in 
mass in the $\Lambda$CDM simulation at $z_1=0$.  Results for other 
$z_1$ and for the SCDM model are qualitatively similar.  
The symbols show that, in all but the bottom right panel, 
$M(r)>1$ on small scales.  To estimate the statistical significance 
of the difference from unity, we have randomized the marks and 
measured $M(r)$, and then repeated this procedure one hundred times.  
The solid curve which is close to unity on all scales shows the 
mean $M(r)$, averaged over these random realizations.  The standard 
deviation around this mean is shown as the dotted line; comparison 
of the symbols with the shape of this curve suggests that we have 
weak evidence that close pairs have larger formation redshifts than 
average.  

\begin{figure}
\centering
\epsfxsize=\hsize\epsffile{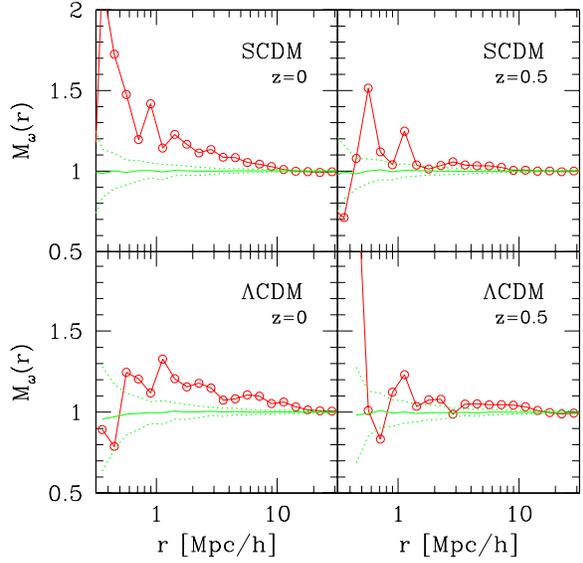}
\caption{Same as previous figure, but now with scaled formation 
redshift as the mark.  All panels show clear evidence for that 
close pairs tend to have larger formation redshifts.  }
\label{xiwf}
\end{figure}

The bottom right panel appears to behave differently from the 
others, so a word on why this happens is in order.  Notice that 
this bin contains a much larger range of halo masses.  Massive haloes 
cluster more strongly than low mass ones, so it is plausible that 
each of the closest pairs in this bin is either a pair of massive 
haloes, or has one of these massive haloes as one member of the pair.  
Since massive haloes have, on average, the smallest formation 
redshifts, $M(r)$ drops on small scales.  This again illustrates 
the complexities of interpretting plots which average over a range 
of masses.  

We have repeated this analysis but using the recaled formation times 
$\omega_{0.5}$ defined in the previous section.  
Since this rescaling removes the dependence of formation redshift on 
mass, we are now allowed to average over the entire range of halo 
masses when computing the marked correlation function, thus 
allowing us to increase the signal-to-noise of our measurement.  
The results are shown in Figure~\ref{xiwf}; now there is clear 
evidence that close pairs tend to form earlier than do more distant 
pairs.  At $1h^{-1}$Mpc, for instance, $M_\omega(r)\sim 1.2$; 
because $M_\omega(r)$ depends on the product of the marks, this 
suggests that haloes with neighbours on this scale tend to have values 
of $\omega_{0.5}$ which are typically $\sim 1.2^{0.5}$ times the 
mean value, but that on scales of order $10h^{-1}$Mpc, this excess 
is considerably smaller.  

The signal appears to be stronger at $z=0$ than at $z=0.5$.  
Some of this is because, as a consequence of the finite mass 
resolution of the simulations, we do not sample haloes with small 
values of $M/M_*$ at higher redshifts.  It will be interesting to 
repeat this analysis on simulations with better mass resolution, so 
as to quantify how the trend with environment depends on $M/M_*$, 
and how it depends on redshift (e.g., one might wonder if the signal 
is clearer at $z=0$ than at $z=0.5$ because the correlation with 
environment shifts to smaller scales at higher redshifts).

\section{Alternative definition of formation}\label{alt}
In the previous sections, formation was defined as the first time 
that at least half the mass of a halo is contained in one progenitor.  
To illustrate that the trend with environment we find does not 
depend on this precise definition of formation, we have modified the 
definition as follows:  here formation is defined as the first time 
that at least half the mass of a halo is contained in progenitors 
each containing at least one percent of the mass of their parent halo.  
The previous definition corresponds to requiring that the minimum 
mass of a progenitor be one-half that of the parent.  
If $z_{0.01}$ denotes this formation time, then $z_{0.01}>z_{0.5}$, 
but the two formation times depend similarly on halo mass:  
in both cases, massive haloes form later.  

\begin{figure}
\centering
\epsfxsize=\hsize\epsffile{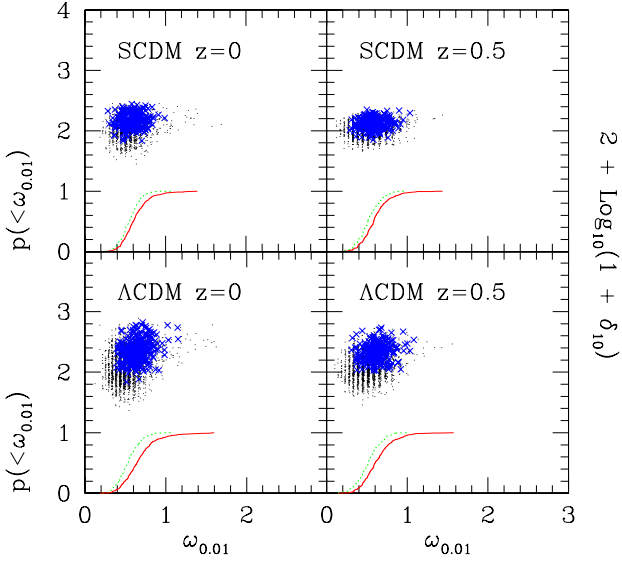}
\caption{Similar to Figure~\ref{wden}, but now with formation 
 defined as the first time that half the final mass is contained 
 in progenitors which are each more massive than one percent of 
 the parent.  Solid and dotted lines show the cumulative 
 formation time distribution for haloes in denser and less dense 
 regions, respectively: Haloes in dense regions tend to have formation 
 times which extend to higher redshifts compared to their counterparts 
 in less dense regions.  }
\label{wnfwden}
\end{figure}

\begin{figure}
\centering
\epsfxsize=\hsize\epsffile{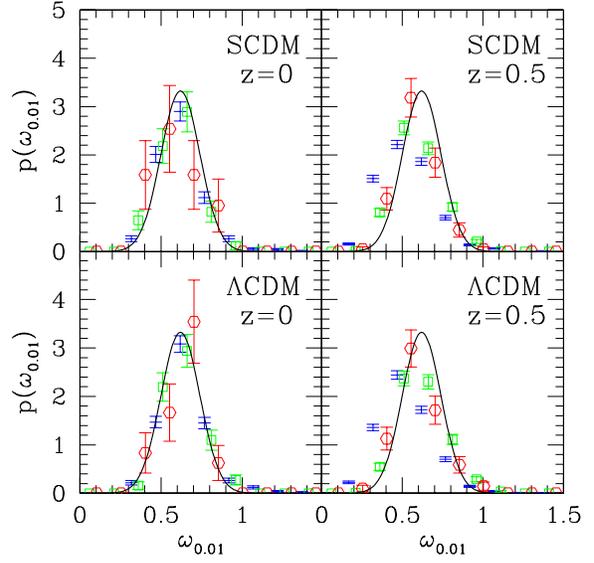}
\caption{Similar to Figure~\ref{pwlc}, but now with formation 
 defined as in Figure~\ref{wnfwden}.  
 In this scaled variable, the distribution of formation 
 times is approximately independent of mass.
 Solid line, drawn to guide the eye, shows a Gaussian 
 distribution with mean 0.62 and rms 0.12.  }
\label{pwnfw}
\end{figure}

Figure~\ref{wnfwden} shows the result of rescaling $z_{0.01}$ 
to $\omega_{0.01}$, similarly to how we rescaled 
$z_{0.5}\to \omega_{0.5}$, and then measuring $p(\omega_{0.01})$ 
in different environments, similarly to what was done in 
Figure~\ref{wden}.  
(Figure~\ref{pwnfw}, the analogue of Figure~\ref{pwlc}, 
shows that this rescaling removes most of the dependence on the 
mass of the parent halo mass and the redshift when it virialized.)
Figure~\ref{wnfwden} shows that $p(<\omega_{0.01})$ is shifted 
towards higher redshifts for the haloes in denser regions 
(solid curve), indicating that for this definition of formation 
also, haloes in dense regions form slightly earlier.  

\begin{figure}
\centering
\epsfxsize=\hsize\epsffile{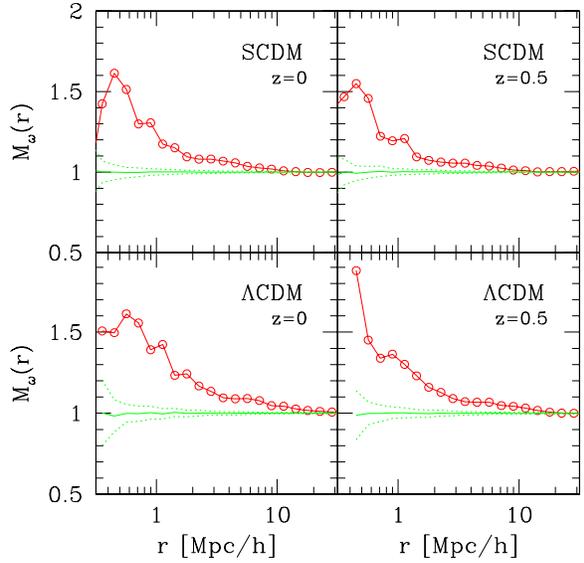}
\caption{Similar to Figure~\ref{xiwf}, but now with formation 
 defined as the first time that half the final mass is contained 
 in progenitors which are each more massive than one percent of 
 the parent.  }
\label{xiwnfw}
\end{figure}

In the previous section, we argued that the marked correlation 
function analysis is a simple and powerful technique for detecting 
and quantifying environmental dependence.  To emphasize this point, 
Figure~\ref{xiwnfw} shows a marked correlation function 
analysis using $\omega_{0.01}$ as the mark.  Notice how similar this 
signal is to that shown in Figure~\ref{xiwf}, in which $\omega_{0.5}$ 
was used as the mark, and also how much easier it is to see the 
environmental dependence here, than in Figure~\ref{wnfwden}.  

Although we will not pursue this further here, we think it is worth 
discussing briefly what information we think the shape of the 
$M(r)$ contains.  
Suppose we write the weighted correlation function as a bias factor 
times the unweighted correlation function:  
 $W(r)\approx B^2(r)\,\xi(r)$.  
In models where environmental effects only matter on small scales, 
one might reasonably expect this bias factor to become independent 
of scale on sufficiently large scales.  Since $\xi(r)$ typically 
decreases with increasing separation, the rate at which 
 $M(r)=[1+B^2\xi(r)]/[1+\xi(r)]\to 1$ 
on large scales is determined by the bias factor.  
This suggests a generalization of the test we have performed:  
smooth the marked field (e.g., compute the mean formation time in 
spheres of radius $R$), compute the correlation function of the 
smoothed marked field, and normalize by the correlation function of 
the smoothed density field.  If the mark is determined by local 
effects, then a measurement of the shape of the marked correlation 
function on large smoothing scales is a measurement of the bias 
associated with the mark.

\begin{figure}
\centering
\epsfxsize=\hsize\epsffile{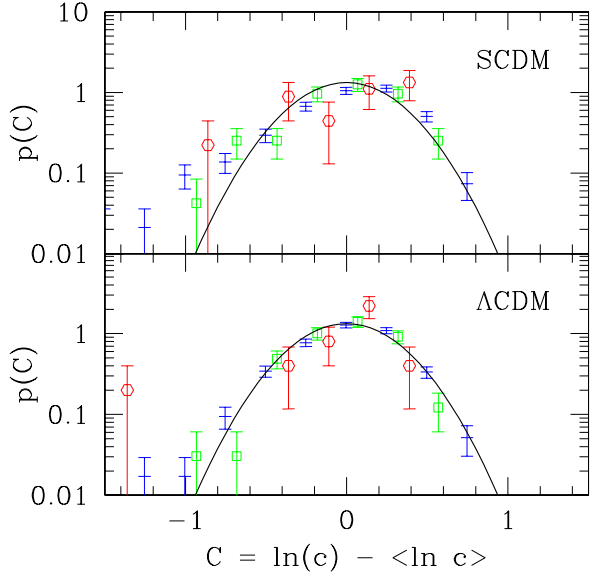}
\caption{Scaled distribution of halo concentrations at $z=0$.  
 Bars, squares and hexagons show results for halos with masses 
 in the range $1< M/M_*\le 2$, $4< M/M_*\le 8$, and $16< M/M_*\le 32$.  
 In this scaled variable, the distribution of halo concentrations 
 is approximately independent of mass.  Solid line shows that this 
 shape is well approximated by a Lognormal with rms 
 $\sigma_{{\rm ln}\,c} = 0.3$.}
\label{plnc}
\end{figure}

\section{Discussion}\label{discuss}
We presented evidence that halo formation is weakly correlated with 
the surrounding density field.  The weakness of the correlation 
suggests that the usual excursion set model neglect of correlations 
is a good approximation, but the fact that a correlation does exist 
means that, as the available data on galaxy properties and the 
surrounding large scale structure becomes more precise, it will 
become necessary to build a more sophisticated model.  

Our demonstration that environmental effects do leave a mark on the 
halo population, and that marked correlation functions are a useful 
method of detecting and quantifying this, suggests that a marked 
correlation analysis of halo concentrations, spins, shapes and 
alignments should yield interesting results.  
For instance, the particular definition of halo formation we chose 
to study in Section~\ref{alt} is motivated by the work of 
Navarro, Frenk \& White (1997) who argue that $z_{0.01}$ correlates 
strongly with the central concentration of the parent halo.  
Since our analysis indicates that halo formation depends on environment, 
it is plausible that the structural properties of haloes will also 
depend on environment.

As a first step, we have attempted a measurement which uses the halo 
concentration as the mark.  This is not entirely straightforward, 
because the GIF simulations we have used in this paper do not have 
particularly good mass-resolution, so they are not well suited to 
estimating the density run around halo centres; as a result, the halo
concentrations we estimate are rather noisy.  Nevertheless, if we 
select a sample for which the Navarro et al. formula is a better 
fit, then we find reasonable agreement with previous higher 
resolution studies (e.g. Jing 2000) which were restricted to 
considerably smaller halo samples.  Specifically, we find that 
on average, massive halos have slightly smaller concentration 
parameters:  $\langle\ln c\rangle = \ln[c_*\,(M/M_*)^{-0.1}]$, 
with $c_*=8$ and 9 at $z=0$ in the SCDM and $\Lambda$CDM runs.  
More importantly, the distribution around the mean concentration is 
approximately independent of halo mass, and is well approximated by 
a lognormal with rms $\sigma_{{\rm ln}\,c}=0.3$.  
This is shown in Figure~\ref{plnc}.  

\begin{figure}
\centering
\epsfxsize=\hsize\epsffile{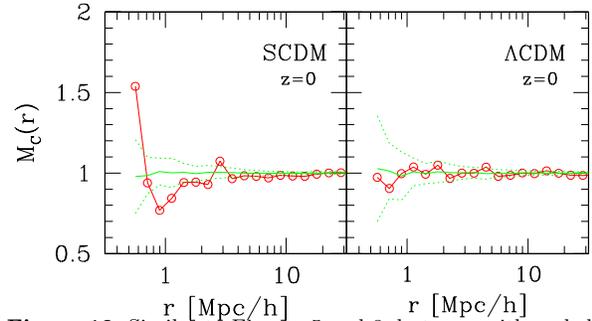}
\vspace{-4cm}
\caption{Similar to Figures~\ref{xiwf} and~\ref{xiwnfw}, but now 
 with scaled concentration $C$ as the mark.  There is no evidence that, 
 once mass dependent trends have been removed, halo concentrations 
 depend on environment.  It will be interesting to see if this 
 conclusion remains when this analysis is repeated with higher 
 resolution simulations.  }
\label{xiconc}
\end{figure}

Because the same lognormal shape provides a good description of the 
distribution of concentrations at all masses, we can use it to remove 
mass-dependent trends, as we did in our study of halo formation in 
main text.  Figure~\ref{xiconc} shows a marked correlation function 
in which this scaled concentration was used as the mark.  There is 
no evidence that the concentrations of close pairs are any different 
than those of more widely separated pairs, suggesting that, once 
mass-dependent trends have been removed, halo concentrations do not 
depend on environment.  Since this strongly suggests that the 
connection between halo formation and concentration is not as 
straightforward as is generally assumed, it will be interesting to 
see if this conclusion remains when the analysis is repeated on 
higher resolution simulations.  

Although we believe a marked correlation analysis of halo 
concentrations, spins, shapes and alignments will yield interesting 
results, we think that there is one measurement which is more 
interesting still.  
Recent studies of pure dark-matter simulations suggest that galaxies 
are likely associated with the substructure components of dark matter 
haloes (Kravtsov et al. 2003).  A marked correlation correlation analysis 
which uses the number of subclumps as the mark would be extremely 
interesting, because one of the crucial assumptions in current 
interpretations of the galaxy correlation function is that the number 
of galaxies which form in a halo depends on halo mass, but does not 
depend on halo environment.  For similar reasons, a marked correlation 
analysis of the sites in which gas cools in dark-matter plus 
hydro- simulations will be very interesting, particularly in view 
of the fact that conventional tests find little evidence for 
environmental trends (Berlind et al. 2003).  

Finally, it is worth mentioning that models based on the work of 
Efstathiou \& Rees (1988) which relate the formation of massive 
black holes to halo formation assume that, at fixed mass, 
the clustering of halos is independent of whether or not they 
formed recently.  Measurements in simulations of the large-scale 
clustering of merger sites indicate that, on scales larger than 
a few Mpc, this assumption is accurate (Percival et al. 2003).
Similarly, simulations show that the large-scale cross-correlation 
between halos that formed recently and the entire halo population 
is similar to the auto-correlation function of the entire halo 
population (Kauffmann \& Haehnelt 2002).  
Both measurements indicate that, at least on large scales, the only 
trends with environment are those which arise from the correlation 
between mass and environment, consistent with the simplest excursion 
set prediction.  

Although they emphasized what they saw on large scales, on smaller 
scales both Kauffmann \& Haehnelt (2002) and Percival et al. (2003) do 
see weak evidence for small additional trends with environment.  Their 
measurements can be reconciled with the excursion set approach if we 
recall that the excursion set prediction is based on the assumption 
that different scales are uncorrelated---in the jargon, this comes 
from using a smoothing filter which is sharp in $k$-space.  
The predictions of the excursion set approach do depend on the choice 
of filter.  However, the precise choice of filter cannot matter on 
scales larger than the correlation length.  Thus, the predicted 
importance of halo mass rather than environment {\em on large scales}, 
although it is derived from consideration of the sharp $k$-space 
filter, is a generic prediction of the approach.  On the other hand, 
one does expect filter-dependent environmental trends on smaller 
scales.  Since the sharp $k$-space filter is not expected to be a 
reasonable choice on small scales, one generically expects to find 
correlations with environment on small scales (over and above those 
associated with halo mass).  Presumably, it is these correlations 
which our analysis is well-suited to detecting.

\bigskip

We would like to thank the Aspen Center for Physics for support, and 
for providing the stimulating environment in which this work was begun.    
We would also like to thank the Virgo consortium for making the 
simulation data used here publically available at   
{\tt http://www.mpa-garching.mpg.de/Virgo}, and Andy Connolly, 
Bob Nichol and the other members of the Pittsburgh Computational 
Astrostatistics (PiCA) Group for providing a fast code with which 
to evaluate marked correlation functions.  
This work was supported by NASA grant NAG5-13270, and by an 
FR Type II Grant from the University of Pittsburgh.


\begin{thebibliography}{99}
\bibitem{abetal} Berlind A. A., Weinberg D. H., Benson A. J., 
                 Baugh C. M., Cole S., Dav\'e R., Frenk C. S., 
                 Jenkins A., Katz N., Cedric G., 2003, ApJ, 593, 1
\bibitem{bcek91} Bond J. R., Cole S., Efstathiou G., Kaiser N.,  1991, ApJ, 
379, 440
\bibitem{er} Efstathiou G., Rees M J., 1988, MNRAS, 230, 5P
\bibitem{ep83} Epstein R., 1983, MNRAS, 205, 207
\bibitem{csf} Frenk C. S., Colberg J. M. Couchman H. M. P., et al., 
  2000, astro-ph/0007362
\bibitem{jing} Jing Y. P., 2000, ApJ, 535, 30
\bibitem{kcdw} Kauffmann G., Colberg J. M., Diaferio A., 
               White S. D. M., 1999, MNRAS, 303, 188
\bibitem{kh}  Kauffmann G., Haehnelt M., 2002, MNRAS, 332, 526
\bibitem{ak03} Kravtsov A., et al. 2003, ApJ, submitted (astro-ph/0308519) 
\bibitem{lc93} Lacey C., Cole S., 1993, MNRAS, 262, 627
\bibitem{lk98} Lemson G., Kauffmann G., 1999, MNRAS, 302, 111
\bibitem{mw96} Mo H. J., White S. D. M., 1996, MNRAS, 282, 347
\bibitem{nfw97} Navarro J., Frenk C., White S. D. M., 1997, ApJ, 490, 493
\bibitem{adi} Nusser A., Sheth R. K., 1999, MNRAS, 303, 685
\bibitem{pjep} Peebles J. E. P., 1993, Principles of Physical Cosmology, 
               Princeton Univ. Press, Princeton, NJ
\bibitem{pspd} Percival W. J., Scott D., Peacock J. A., Dunlop J. S., 
               2003, MNRAS, 338, 31P
\bibitem{st99} Sheth R. K., Tormen G., 1999, MNRAS, 308, 119
\bibitem{smt99} Sheth R. K., Mo H. J., Tormen G., 2001, MNRAS, 323, 1
\bibitem{st02} Sheth R. K., Tormen G., 2002, MNRAS, 329, 61
\bibitem{st04} Sheth R. K., Tormen G., 2004, MNRAS, in press (astro-ph/0402055)
\bibitem{w96} White S. D. M., 1996, in Cosmology and Large Scale
  Structure, Les Houches Session LX, eds. R. Schaefer, J. Silk, 
  M. Spiro, and J. Zinn-Justin, Elsevier, Amsterdam, p. 349
\end{thebibliography}
\end{document}